\documentclass[a4paper]{article}

\usepackage{INTERSPEECH2018}
\usepackage{graphicx}
\usepackage{subcaption}
\usepackage{xcolor}
\usepackage{colortbl}
\usepackage{multirow}
\usepackage{tabularx}
\usepackage{enumitem}

\graphicspath{{./}{figures/}}
\def\x{{\mathbf x}}
\def\h{{\mathbf h}}
\newcolumntype{V}{>{\centering\arraybackslash} m{.2\linewidth} }

\title{Triplet Network with Attention for Speaker Diarization}
\name{Huan Song$^1$ \thanks{This work was supported in part by the SenSIP center at Arizona State University. This work was performed under the auspices of the U.S. Dept. of Energy by Lawrence Livermore National Laboratory under Contract DE-AC52- 07NA27344.}, Megan Willi$^2$, Jayaraman J. Thiagarajan$^3$, Visar Berisha$^{1,2}$, Andreas Spanias$^1$}
\address{
  $^1$SenSIP Center, School of ECEE, Arizona State University, Tempe, AZ\\
  $^2$Department of Speech and Hearing Science, Arizona State University, Tempe, AZ\\
  $^3$Lawrence Livermore National Labs, 7000 East Avenue, Livermore, CA}
\email{\{huan.song, megan.willi, visar, spanias\}@asu.edu, jjayaram@llnl.gov}

\begin{document}

\maketitle
\begin{abstract}
In automatic speech processing systems, speaker diarization is a crucial front-end component to separate segments from different speakers. Inspired by the recent success of deep neural networks (DNNs) in semantic inferencing, triplet loss-based architectures have been successfully used for this problem. However, existing work utilizes conventional i-vectors as the input representation and builds simple fully connected networks for metric learning, thus not fully leveraging the modeling power of DNN architectures. This paper investigates the importance of learning effective representations from the sequences directly in metric learning pipelines for speaker diarization. More specifically, we propose to employ attention models to learn embeddings and the metric jointly in an end-to-end fashion. Experiments are conducted on the CALLHOME conversational speech corpus. The diarization results demonstrate that, besides providing a unified model, the proposed approach achieves improved performance when compared against existing approaches.

\end{abstract}

\section{Introduction}
With the ever-increasing volume of multimedia content on the Internet, there is a crucial need for tools that can automatically index and organize the content. In particular, speaker diarization deals with the problem of indexing speakers in a collection of recordings, without \textit{a priori} knowledge about the speaker identities. In scenarios where the single-speaker assumption of recognition systems is violated, it is critical to first separate speech segments from different speakers prior to downstream processing. Typical challenges in speaker diarization include the need to deal with similarities between a large set of speakers, differences in acoustic conditions, and the adaptation of a trained system to new speaker sets. 

An important class of diarization approaches rely on extracting i-vectors to represent speech segments, and then scoring similarities between i-vectors using pre-defined similarity metrics (e.g. cosine distance) to achieve speaker discrimination. Despite its widespread use, it is well known that the i-vector extraction process requires extensive training of a Gaussian Mixture Model based Universal Background Model (GMM-UBM) and estimation of the total variability matrix (i-vector extractor) beforehand using large corpora of speech recordings. While several choices for the similarity metric currently exist, likelihood ratios obtained through a separately trained Probabilistic Linear Discriminant Analysis (PLDA) model are commonly utilized \cite{Khoury_ODYSSEY_2014}.

More recently, with the advent of modern representation learning paradigms, designing effective metrics for comparing i-vectors has become an active research direction. In particular, inspired by its success in computer vision tasks \cite{schroff2015facenet, lin2015learning, hoffer2015deep}, many recent efforts formulate the diarization problem as deep metric learning \cite{le2017triplet, garcia2017speaker, bredin2017tristounet}. For instance, a triplet network that builds latent spaces, wherein a simple Euclidean distance metric is highly effective at separating different classes, is a widely adopted architecture. However, in contrast to its application in vision tasks, metric learning is carried out on the i-vector representations instead of the raw data \cite{le2017triplet}. Consequently, the first stage of the diarization pipeline stays intact, while the second stage is restricted to using fully connected networks. Though this modification produced state-of-the-art results in diarization and outperformed conventional scoring strategies, it does not support joint representation and task-based learning, which has become the \textit{modus operandi} in deep learning. On the other hand, Garcia-Romero et al. \cite{garcia2017speaker} propose to perform joint embedding and metric learning, but use siamese networks for metric learning, which have generally shown poorer performance when compared to triplet networks \cite{hoffer2015deep}.

In this paper, we propose to explore the use of joint representation learning and similarity metric learning with triplet loss in speaker diarization, while entirely dispensing the need for i-vector extraction. Encouraged by the recent success of \textit{self-attention} mechanism in sequence modeling tasks \cite{vaswani2017attention, song2017attend}, for the first time, we leverage attention networks to model the temporal characteristics of speech segments. Experimental results on the CALLHOME corpus demonstrate that, with an appropriate embedding architecture, triplet network applied on raw audio features from a comparatively smaller dataset outperforms the same applied on i-vectors, wherein the GMM-UBM was trained using a much larger corpus.
\section{Related Work}
In this section, we briefly review the recent literature on techniques for speaker diarization. Over the last few years, speaker diarization approaches have quickly evolved from the traditional MFCC based GMM segmentation and BIC clustering \cite{tranter2006overview,anguera2012speaker,barras2006multistage} to systems centered around i-vector representations \cite{prazak2011speaker, sell2014speaker}. Initially proposed for speaker verification tasks \cite{dehak2011front}, i-vectors are low-dimensional features extracted over variable-length speech segments to compensate for within and between-speaker variabilities. Different speakers can then be effectively discriminated by utilizing either standard similarity metrics (e.g. cosine distance) \cite{shapiro2015clustering} or likelihood ratios from PLDA \cite{kenny2010bayesian}) to cluster i-vectors from the segments. 

More recently, several deep learning-based solutions have been developed to automatically infer similarity metrics to compare speech segments. More specifically, supervised metric learning architectures namely \textit{siamese} \cite{chopra2005learning, koch2015siamese} and \textit{triplet} \cite{hoffer2015deep, schroff2015facenet} networks are prevalent. Broadly speaking, these architectures infer a non-linear mapping $\mathcal{A}(\cdot)$, such that, in the resulting latent space the within-class sample distances are minimized while the between-class distances are maximized based on a certain margin. For instance, in \cite{le2017triplet}, Lan \textit{et al.} proposed to employ triplet networks on i-vectors to infer a similarity metric, and achieved state-of-the-art results over conventional metrics in the diarization literature. Despite its effectiveness, it is important to note that the feature extraction process is disentangled from the metric learning network and hence cannot support end-to-end inferencing. However, recent success of such end-to-end learning systems in computer vision applications \cite{krizhevsky2012imagenet, gordo2017end, song2018optimizing} motivates the design of a deep metric learning architecture that works directly on the temporal sequences.  

Long Short-Term Memory (LSTM) based recurrent networks have become the \textit{de facto} solution to sequence modeling tasks including acoustic modeling \cite{sak2014long}, speech recognition \cite{graves2013speech} and Natural Language Processing (NLP) \cite{sundermeyer2012lstm}. Recently, architectures entirely based on attention mechanism have shown promising value in sequence-to-sequence learning \cite{vaswani2017attention} and clinical data analysis \cite{song2017attend}. Besides providing significantly faster training, attention networks demonstrate efficient modeling of long-term dependencies.

In this paper, we utilize attention networks with the triplet ranking loss to jointly learn embeddings and a similarity metrics for speech segments. To the best of our knowledge, the approaches in \cite{bredin2017tristounet} and \cite{garcia2017speaker} are the most related to our work. While Bredin \textit{et al.} \cite{bredin2017tristounet} used triplet networks based on LSTMs, they applied it to a simpler binary classification task of speaker turn identification. Whereas, in \cite{garcia2017speaker}, Romero \textit{et al.} performed a similar joint learning for diarization, but based on a siamese network. Compared to the triplet ranking loss, which requires a margin to be satisfied for each given reference sample, the cross-entropy loss used in \cite{garcia2017speaker} requests correct prediction of all different-speaker or same-speaker pairs and hence exhibits much less flexibility. 

%We suspect this to be a reason for the observation that the approach in \cite{garcia2017speaker} could not significantly outperform the PLDA baseline.

% Specifically, we demonstrate that, using the carefully designed network architecture, better diarization performance can be achieved using a single DNN model and much less training efforts (as shown by the comparison in Figure \ref{fig:arch}). 

%There are different types of network architectures that have provisions for implicit learning of the data representations, computing the similarity between the inputs, ranking and classification \cite{berlemont2015siamese,hoffer2015deep}. One such architecture is the Siamese Network which has been shown to perform well for classification tasks \cite{berlemont2015siamese}. However these networks are prone to global and contextual classification issues as reported in \cite{hoffer2015deep}. A novel architecture inspired from the Siamese Network namely the Triplet Network is proposed in \cite{hoffer2015deep} which comprises of three instances of a shared parameter feedforward network and is expected to provide better embedding models. Some of the other architectures derived from the Siamese network perform well for different applications in other domains such as image localization, classification \cite{lin2015learning}and most importantly in speaker diarization \cite{garcia2017speaker}. 
\section{Proposed Approach}

\begin{figure}[t]
    \centering
    \begin{subfigure}[b]{1\linewidth}
        \centering
        \includegraphics[width=1\linewidth]{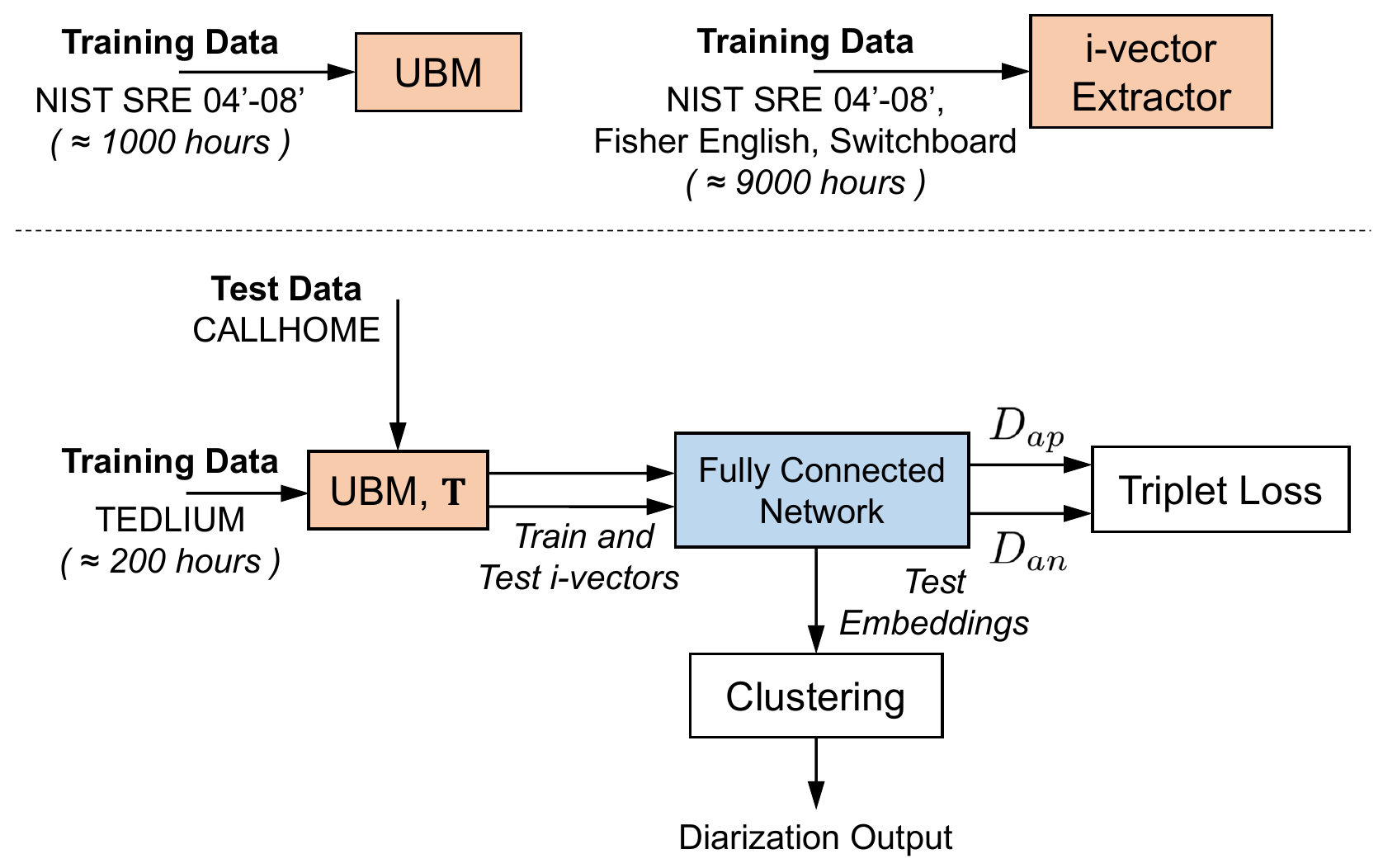}
        \caption{Baseline approach: (top) i-vectors are extracted using a large corpus of recordings with GMM-UBM and i-vector extractor modules; (bottom) Similarity metric is trained using a triple network trained on the i-vectors.}
    \end{subfigure}%
    \\
    \begin{subfigure}[b]{1\linewidth}
        \centering
        \includegraphics[width=0.75\linewidth]{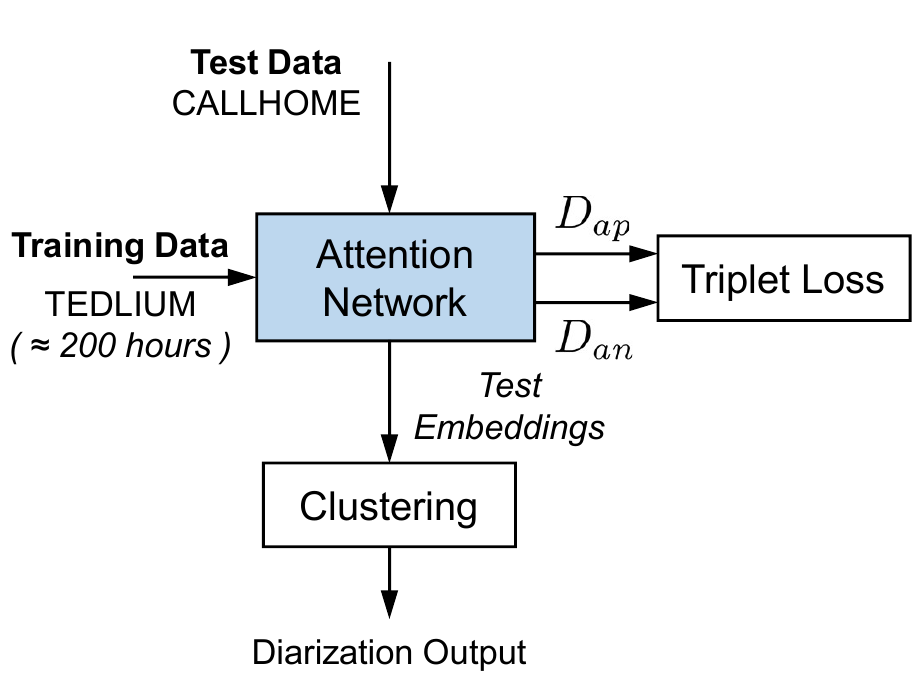}
        \caption{Proposed approach: Joint learning of embedding and similarity metric for diarization. As observed, it completely eliminates the i-vector extraction process and enables effective training with limited data.}
    \end{subfigure}
    \caption{Comparison of diarization strategies and training data requirements for the baseline approach in \cite{le2017triplet} and the proposed approach.}
    \label{fig:arch}
 \vspace{-0.2in}
\end{figure}

As shown in Figure \ref{fig:arch}(b), the proposed approach works directly with raw temporal speech features to learn a similarity metric for diarization. Compared to the baseline in Figure \ref{fig:arch}(a), the two-stage training process is simplified into a single end-to-end learning strategy, wherein deep attention models are used for embedding computation and the triplet loss is used to infer the metric. Similar to existing diarization paradigms, we first train our network using out-of-domain labeled corpus, and then perform diarization on a target dataset using unsupervised clustering. In the rest of this section, we describe the proposed approach in detail.

\subsection{Temporal Segmentation and Feature Extraction}
\label{sec:preprocessing}

For the speech recordings, we first perform non-overlapping temporal segmentation into $2$-second segments. Following the Voice Biometry Standardization (VBS) \footnote{http://voicebiometry.org/}, we extract MFCC features using $25$ms Hamming windows with $15$ms overlap. After adding delta and double-delta coefficients, we obtain $60$-dimensional feature vectors at every frame. Consequently, each data sample corresponds to a temporal sequence feature $\x_i\in \mathbb{R}^{T\times d}$, where $T$ is the number of frames in each segment and $d=60$ is the feature dimension.

\begin{figure}[t]
	\centering
	\includegraphics[width=\linewidth]{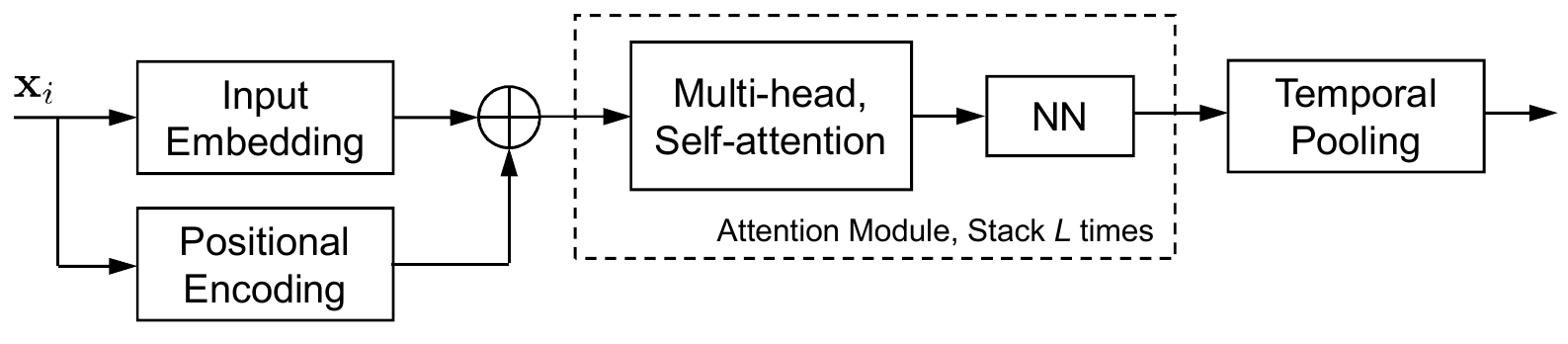}
	\caption{Illustration of the attention model used for computing embeddings from MFCC features of speech segments.}
	\label{fig:attention}
\vspace{-0.2in}
\end{figure} 

\subsection{Embeddings using Attention Models}
\label{sec:attention}
As described earlier, we use attention models to learn embeddings directly from MFCC features for the subsequent metric learning task. The attention model used in our architecture is illustrated in Figure \ref{fig:attention}. The module comprised of a multi-head, self-attention mechanism is the core component of the attention model \cite{vaswani2017attention}. More specifically, denoting the input representation at layer $\ell$ as $\{\h_t^{\ell-1}\}_{t=1}^T$, we can obtain the hidden representation at time step $i$ based on attention as follows:
\begin{gather}
	\h_i^{\ell}=\sum_{t=1}^T w_t^{(i)} \h_t^{\ell-1},~~1\leq i \leq T, \label{eq:att_rep} \\
	w_t^{(i)}=\text{softmax}\left(\frac{\h_i^{\ell-1}\cdot \h_t^{\ell-1}}{\sqrt{D}}\right), \label{eq:att_weight}\\
	\h_i^{\ell} \leftarrow \mathcal{F}(\h_i^{\ell}),
\end{gather}Here, $D=256$ refers to the size of the hidden layer and $\mathcal{F}$ denotes a feed-forward neural network. The attention weight in equation (\ref{eq:att_weight}) denotes the interaction between temporal positions $i$ and $t$ by performing scaled inner product between the two representations. During the computation of hidden representation at time step $i$, $w_t^{(i)}$ weights the contribution from other temporal positions. Note that, these representations are processed by $\mathcal{F}$ before connecting to the next attention module, as shown in Figure \ref{fig:attention}. We employ a $1$D convolutional layer (kernel size is $1$) with ReLU activation \cite{nair2010rectified} for $\mathcal{F}$. Finally, the attention module is stacked $L$ times to learn increasingly deeper representations.

Attention-based representations in equation (\ref{eq:att_rep}) are computed within each speech segment independently and hence this process is referred to as \textit{self-attention}. Furthermore, the hidden representations $\h_t^{\ell}$ are computed using $H$ different network parameterizations, denoted as heads \cite{vaswani2017attention}, and the resulting $H$ attention representations are concatenated together. This can be loosely interpreted as an ensemble of representations. Such a \textit{multi-head} operation facilitates dramatically different temporal parameterizations and significantly expands the modeling power. Our current implementation sets $L=2$ and $H=8$.

Although attention computation explicitly models the temporal interactions, it does not encode the crucial ordering information contained in speech. The front-end positional encoding block handles this problem by mapping every relative frame position $t$ in the segment to fixed locations in a random lookup table. As shown in Figure \ref{fig:attention}, the encoded representation is subsequently added up with the input embedding (obtained also from a $1$D CNN layer). Finally, we include a temporal pooling layer to reduce the final representation $\h^{L} \in \mathbb{R}^{T\times D}$ into a $D$-dimensional vector by averaging along the time-axis.

\subsection{Metric Learning with Triplet Loss}
\label{sec:triplet}
The representations from the deep attention model are then used to learn a similarity metric with the triplet ranking loss. Note that the attention model parameters and the metric learner are optimized jointly using back-propagation. In a triplet network, each input is constructed as a set of $3$ samples $\x=\{\x_p, \x_r, \x_n\}$, where $\x_r$ denotes an anchor, $\x_p$ denotes a positive sample belonging to the same class as $\x_r$, and $\x_n$ a negative sample from a different class. Each of the samples in $\x$ are processed using the attention model (Section 3.2) $\mathcal{A}(\cdot):\mathbb{R}^{T\times d}\mapsto \mathbb{R}^D$ and distances are computed in the resulting latent spaces:
\begin{gather*}
D_{rp}=\lVert \mathcal{A}(\x_r) - \mathcal{A}(\x_p) \rVert_2 \\
D_{rn}=\lVert \mathcal{A}(\x_r) - \mathcal{A}(\x_n) \rVert_2 
\end{gather*}
The triplet loss is defined as
\begin{equation}
\label{eq:triplet}
l(\x_p, \x_r, \x_n)=\max(0, D_{rp}^2-D_{rn}^2+\alpha)
\end{equation}where $\alpha$ is the margin and the objective is to achieve $D_{rn}^2\geq D_{rp}^2+\alpha$. In comparison, the contrastive loss often used in siamese network includes the hinge term $\max(0, \alpha-D_{ij})$ for different-class samples $\x_i$ and $\x_j$, and hence requires $\alpha$ to be a global margin. Such a formulation significantly restricts the model flexibility and expressive power.
%used in \cite{garcia2017speaker} 

Given a large number of samples $N$, the computation of equation (\ref{eq:triplet}) is infeasible among the $\mathcal{O}(N^3)$ triplet space. It is tempting to greedily select the most effective triplets, which maximizes $D_{rp}$ and minimizes $D_{rn}$. Instead of performing such hard sampling, we follow \cite{schroff2015facenet} to sample all possible $\x_p$ and only selecting \textit{semi-hard} $\x_n$: the negative samples satisfying $D_{rp}^2\leq D_{rn}^2\leq D_{rp}^2+\alpha$. Additionally, we adopt an online sampling strategy that restricts the sampling space to the current mini-batch during training. All sampled triplets are gathered to compute the loss in equation (\ref{eq:triplet}). 

For the online sampling scheme, the mini-batch construction step is crucial. Ideally, each batch should cover both a large number of speakers and sufficient samples per speaker. However, we are constrained by the GPU memory ($8$GB) and only able to set maximum batch size $B=256$. We preset $M$ as the number of speakers per batch and when sampling each mini-batch, $M$ speakers are first sampled and $B/M$ speech segments are then sampled for every speaker. As a result, the parameter $M$ represents the trade-off between modeling more speakers each time, and covering sufficient samples for those speakers. In our experiments, $M$ was tuned based on the performance on the development set, as will be discussed in Section \ref{sec:exp:training}.
\section{Experiments}
In this section, we discuss the training process for our approach and evaluate its performance on the CALLHOME corpus.

\begin{figure}[t]
    \centering
    \begin{subfigure}[b]{0.8\linewidth}
        \centering
        \includegraphics[width=0.9\linewidth]{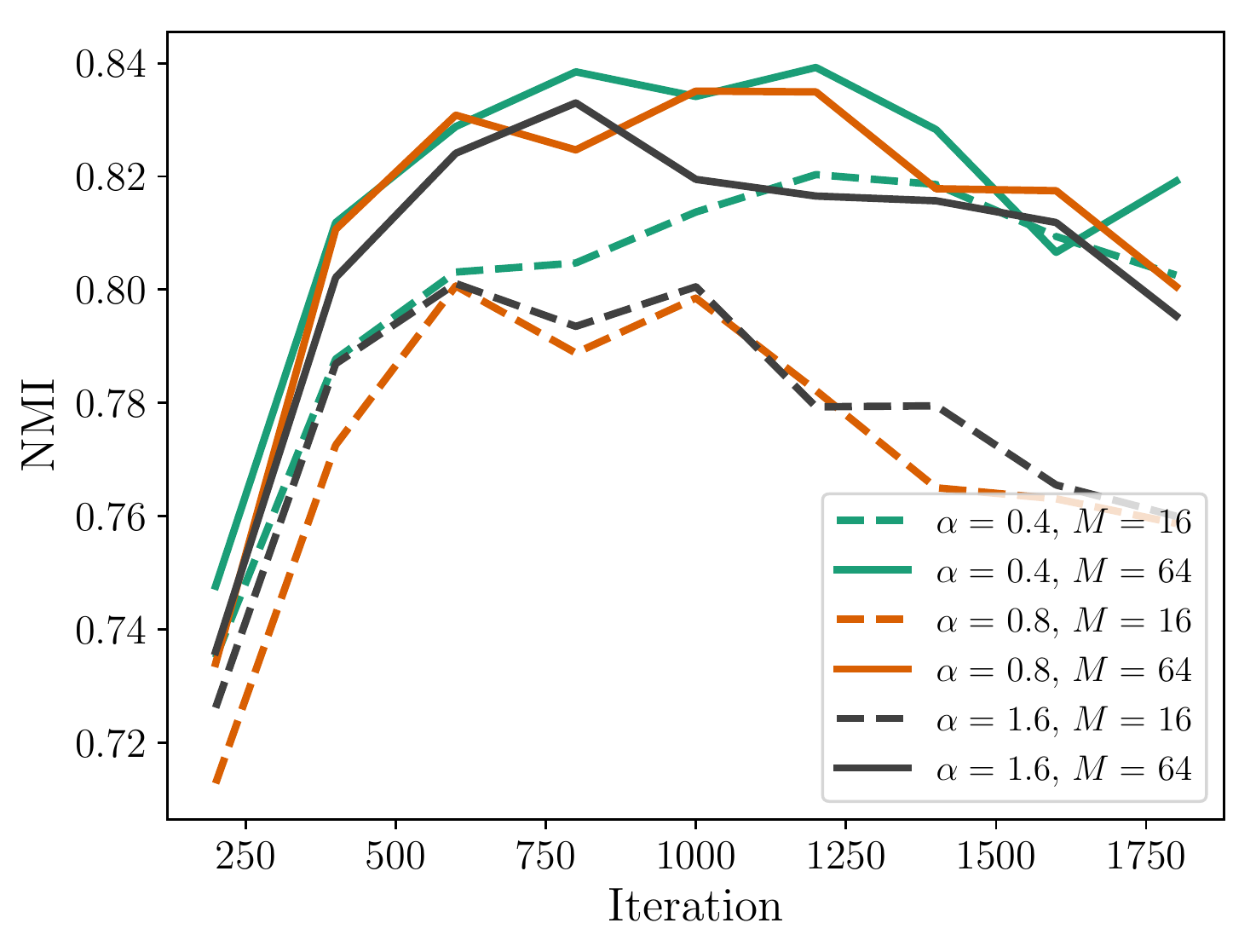}
        \caption{NMI score from the speaker clustering results.}
    \end{subfigure}%
    \\
    \begin{subfigure}[b]{0.8\linewidth}
        \centering
        \includegraphics[width=0.9\linewidth]{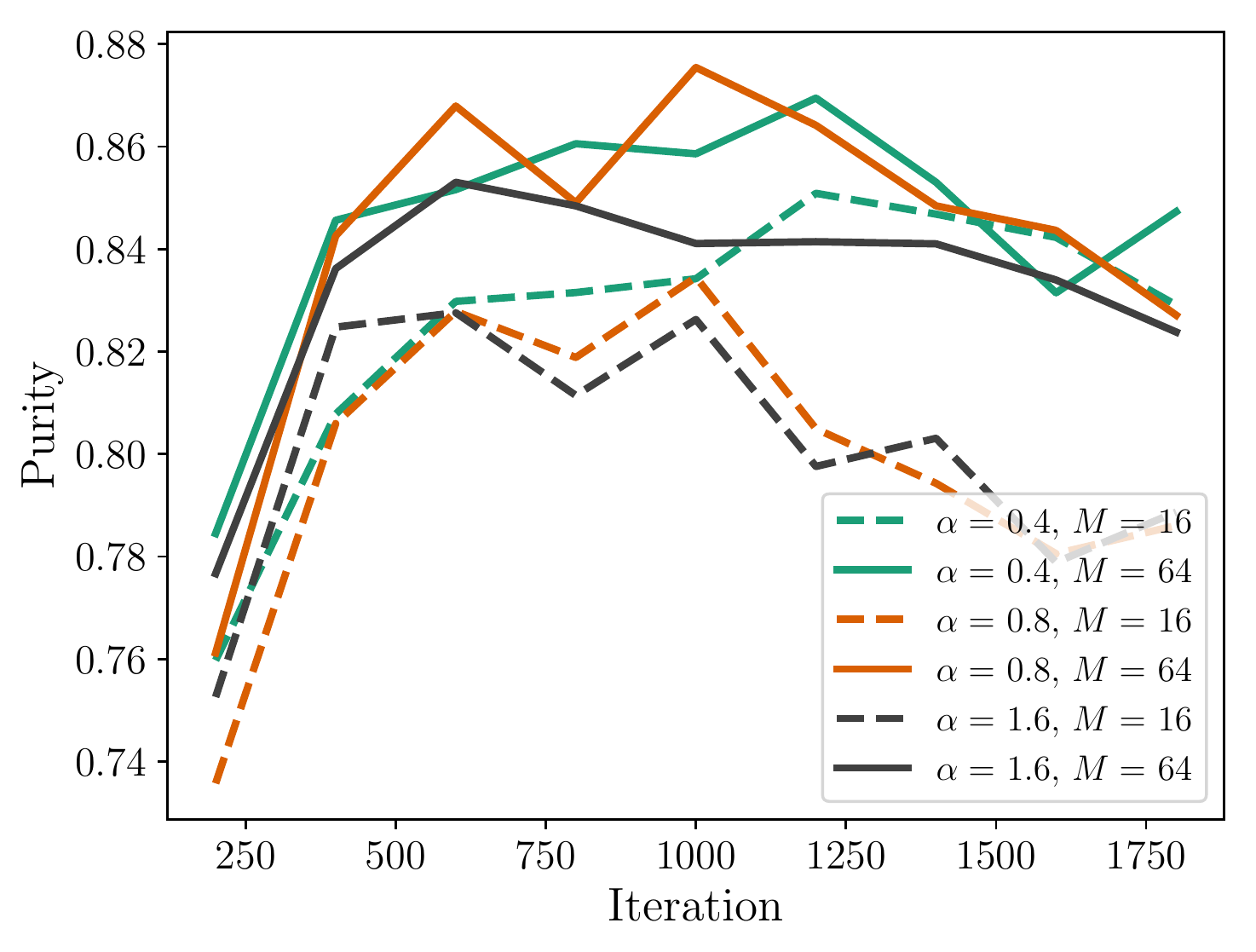}
        \caption{Purity score from the speaker clustering results.}
    \end{subfigure}
    \caption{Parameter tuning on TEDLIUM development set for triplet margin $\alpha$ and number of speakers per batch $M$. Curves for $M=8$ and $32$ are omitted for clarity.}
    \label{fig:param_tuning}
 \vspace{-0.2in}
\end{figure}

\subsection{Triplet Network Training}
\label{sec:exp:training}
The proposed model was trained on the TEDLIUM corpus which consists of $1495$ audio recordings. After ignoring speakers with less than $45$ transcribed segments, we have a set of $1211$ speakers with an average recording length of $10.2$ minutes. All recordings were down-sampled to $8$kHz to match the target CALLHOME corpus. The temporal segmentation and MFCC extraction were carried out as discussed in Section \ref{sec:preprocessing}.

For the proposed approach, there are two important training parameters that need to be selected, i.e.\ triplet margin $\alpha$ and the number of speakers per mini-batch $M$. In order to quickly configure the parameters, we build a training subset by randomly selecting $20\%$ of the total recordings and a development set by taking $50$ recordings from the original TEDLIUM train, dev and test sets. At every $200$ iterations of training on the subset, we extract the embeddings for the development set and perform speaker clustering using $k$-Means, with a known number of speakers. The clustering performance is evaluated by the standard Normalized Mutual Information (NMI) and Purity scores. Based on this procedure, we jointly tuned both parameters by performing a grid search on $\alpha=[0.4, 0.8, 1.6]$ and $M=[8, 16, 32, 64]$. As shown in Figure \ref{fig:param_tuning}, having a higher $M$ value consistently provides better clustering results and alleviates model overfitting. Additionally, a lower triplet margin generally helps the training process. Based on these observations, we configured $\alpha=0.8, M=64$ to train our model on the entire TEDLIUM corpus.

\begin{figure}[t]
	\centering
	\includegraphics[width=0.8\linewidth]{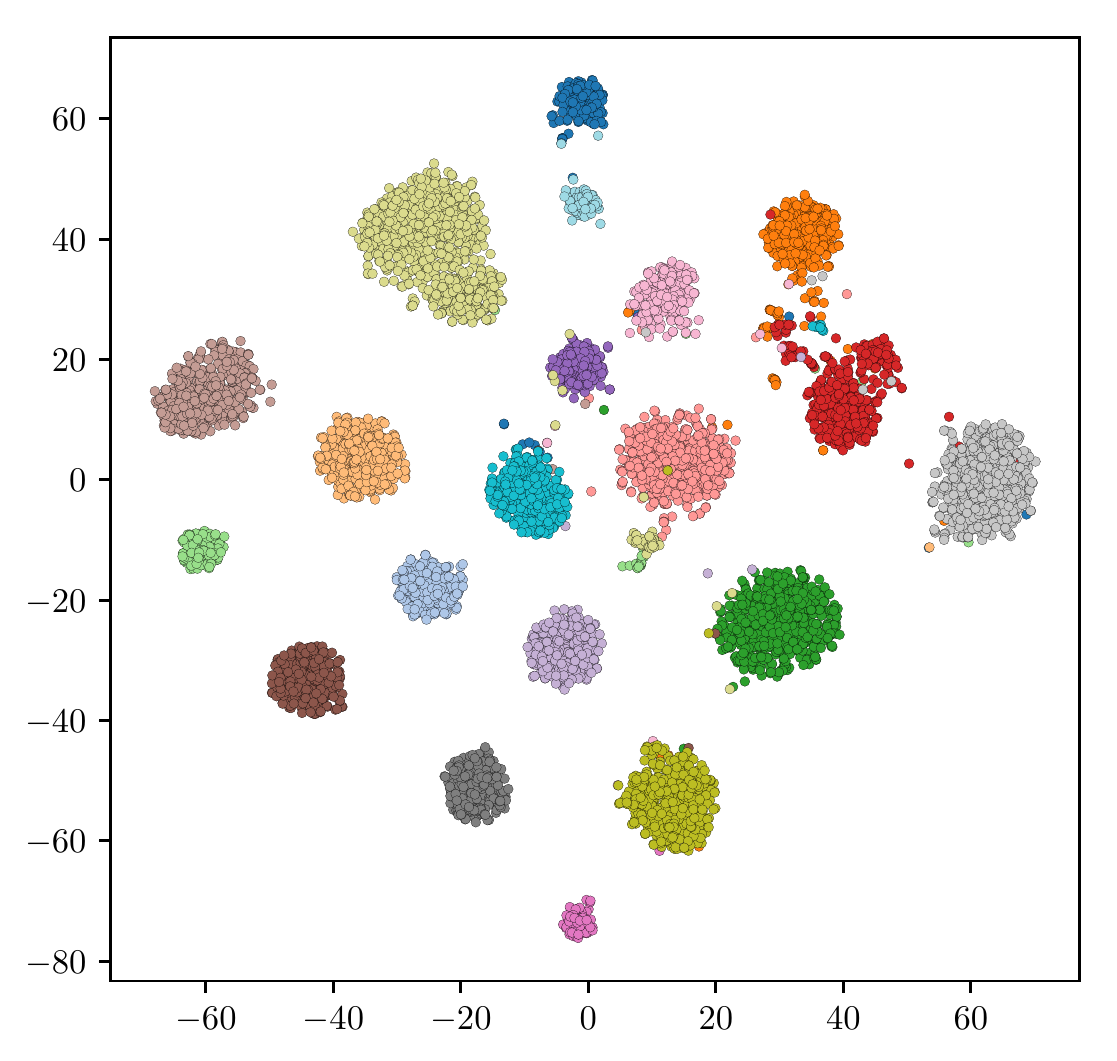}
	\caption{$2$D t-SNE visualization of the first $20$ speakers from TEDLIUM development set. Each point corresponds to one speech segment and they are color coded by the speaker.}
	\label{fig:tsne}
\vspace{-0.2in}
\end{figure}

To study the embeddings from the attention model and the impact of triplet loss, we show the $2$D t-SNE visualization \cite{van2014accelerating} of samples in the development set in Figure \ref{fig:tsne}. It is observed that the model is highly effective at separating unseen speakers and provides little distinction on segments from the same speakers. These embeddings achieve $0.94$ score on both NMI and Purity, with $k$-Means clustering for the development set.

\subsection{Diarization Results}
The trained model is evaluated on the CALLHOME corpus \footnote{https://ca.talkbank.org/access/CallHome/} for diarization performance. CALLHOME consists of telephone conversations in $6$ languages: Arabic, Chinese, English, German, Japanese and Spanish. In total, there are $780$ transcribed conversations containing $2$ to $7$ speakers. After obtaining the embeddings through the proposed approach, we perform x-means \cite{pelleg2000x} to estimate the number of speakers and then use $k$-means clustering with the estimation. We force x-means to split at least $2$ clusters by initializing it with $2$ centroids. Note that there are usually multiple moving parts on complete diarization systems in the literature. In particular, more sophisticated clustering algorithms \cite{wang2017speaker}, overlapping test segments and calibration \cite{sell2014speaker} can be incorporated to improve the overall diarization performance. However, in this work we focus on investigating the efficacy of the DNN modeling and fix the other components in their basic configurations.

We utilize \texttt{pyannote.metric} \cite{bredin2017pyannote} to calculate Diarization Error Rate (DER) as the evaluation metric. Although DER collectively considers false alarms, missed detections and confusion errors, most existing systems evaluated on CALLHOME \cite{sell2014speaker, garcia2017speaker} accounts for only the confusion rate and ignores overlapping segments. Following this convention, we use the oracle speech activity regions and use only the non-overlapping sections. Additionally, there is a collar tolerance of $250$ms at both beginning and end of each segment. We compare the proposed approach with the following baseline systems:

\textbf{Baseline 1: i-vector + cosine/PLDA scoring}. We utilize VBS pre-trained models for i-vector extraction on CALLHOME corpus. The specific GMM-UBM and i-vector extractor training data are shown in Figure \ref{fig:arch}(a). Though different from ours, the training corpus is significantly more comprehensive than the TEDLIUM set we used. The GMM-UBM consists of $2048$ Gaussian components and the i-vectors are $600-$dimensional. We also used the backend LDA model contained in VBS for i-vector pre-processing. In the actual clustering, cosine or PLDA scores are used to calculate the sample-to-centroid similarities at each iteration. 

\textbf{Baseline 2: i-vector + triplet with FCN training}. This baseline is very similar to \cite{le2017triplet} except for $2$ modifications: 1) We do not consider the speaker linking procedure as there are very few repeated speakers in CALLHOME. 2) We use a larger FCN network than \cite{le2017triplet} to allow a fair comparison to the proposed approach. The hidden layers have size $512-1024-512-256$ and batch normalization \cite{ioffe2015batch} is applied at each layer after the ReLU activation. Further, i-vectors are extracted on TEDLIUM based on the transcribed speech sections with average length of $8.6$ seconds. The triplet network is tuned in a similar procedure as in Section \ref{sec:exp:training} and the best parameters were found to be $\alpha=0.4, M=16$.

The comparison between the proposed approach and the baselines is shown in Table \ref{tab:der}. It is observed that baseline $2$ indeed exceeds both conventional i-vector scoring methods. However, our unified learning approach trained on a much smaller TEDLIUM corpus achieves better performance, this evidencing the effectiveness of end-to-end learning.

\begin{table}[t]
  \caption{Diarization Results on CALLHOME Corpus.}
  \label{tab:der}
  \centering
  \begin{tabular}{ V | c | V }
    \hline
    \multicolumn{2}{c|}{\textbf{System}} & \textbf{DER (\%)} \\
    \hline
    \hline
    \multirow{3}{*}{i-vector}	& cosine	& $18.7$	\\
    							& PLDA \cite{Khoury_ODYSSEY_2014}& $17.6$\\
    							& Triplet with FCN \cite{le2017triplet}	& $13.4$	\\
    \hline
    \multicolumn{2}{c|}{\textbf{Proposed Approach}} 		& $\mathbf{12.7}$   \\
    \hline
  \end{tabular}
  \vspace{-0.2in}
\end{table}

%\begin{tabular}{ c | c | c}
%  \hline
%  \multicolumn{2}{c|}{\textbf{System}} & \textbf{DER (\%)} \\ [3ex]
%  \hline
%  \hline
%  \multirow{3}{*}{i-vector}	& cosine	& $18.7$	\\
%    					& PLDA		& $17.6$	\\
%    					& Triplet with FCN	& $13.4$	\\
%  \hline
%  \multicolumn{2}{c|}{Proposed Approach} 		& $12.7$   \\ [3ex]
%  \hline
%\end{tabular}
\section{Conclusions}
This paper studies the role of learning embeddings under a triplet ranking loss for speaker diarization. Results on the CALLHOME corpus show that when compared to training a UBM model and then a separate triplet DNN, the two steps can be combined together to achieve improved performance with less training effort. Future work will investigate more sophisticated sampling strategies for metric learning \cite{manmatha2017sampling} and comparative studies with existing DNN architectures \cite{wang2017speaker, garcia2017speaker}.
%including LSTM \cite{wang2017speaker} and time-delay network \cite{garcia2017speaker}.

\bibliographystyle{IEEEtran}

\bibliography{mybib}

% \begin{thebibliography}{9}
% \bibitem[1]{Davis80-COP}
%   S.\ B.\ Davis and P.\ Mermelstein,
%   ``Comparison of parametric representation for monosyllabic word recognition in continuously spoken sentences,''
%   \textit{IEEE Transactions on Acoustics, Speech and Signal Processing}, vol.~28, no.~4, pp.~357--366, 1980.
% \bibitem[2]{Rabiner89-ATO}
%   L.\ R.\ Rabiner,
%   ``A tutorial on hidden Markov models and selected applications in speech recognition,''
%   \textit{Proceedings of the IEEE}, vol.~77, no.~2, pp.~257-286, 1989.
% \bibitem[3]{Hastie09-TEO}
%   T.\ Hastie, R.\ Tibshirani, and J.\ Friedman,
%   \textit{The Elements of Statistical Learning -- Data Mining, Inference, and Prediction}.
%   New York: Springer, 2009.
% \bibitem[4]{YourName17-XXX}
%   F.\ Lastname1, F.\ Lastname2, and F.\ Lastname3,
%   ``Title of your INTERSPEECH 2018 publication,''
%   in \textit{Interspeech 2018 -- 19\textsuperscript{th} Annual Conference of the International Speech Communication Association, September 2-6, Hyderabad, India Proceedings, Proceedings}, 2018, pp.~100--104.
% \end{thebibliography}

\end{document}